\documentclass[10pt,twocolumn,prl,aps,floatfix,superscriptaddress,longbibliography]{revtex4-1}

\usepackage{color}
\usepackage{graphicx}
\usepackage{amsthm}
\usepackage{amsmath}
\usepackage{amssymb}
\usepackage{enumerate}
\usepackage{placeins}
\usepackage{siunitx}
\usepackage{bibunits}

\DeclareMathOperator{\Tr}{Tr}
\newcommand{\ket}[1]{\vert #1 \rangle}
\newcommand{\eket}[1]{\bigl \vert #1 \bigr \rangle}
\newcommand{\R}{\boldsymbol{R}}

\newcommand{\ebra}[1]{\bigl \langle #1 \bigr \vert}

\newcommand{\figref}[1]{Figure~\ref{#1}}
\renewcommand{\vec}[1]{\boldsymbol{#1}}
\newcommand{\ren}{R\'{e}nyi~}
\newcommand{\Eqref}[1]{Eq.~\eqref{#1}}

\begin{document}

\begin{bibunit}[apsrev4-1]

\title{Entanglement area law  in superfluid $^4$He}

\author{C. M. Herdman}
\email{cherdman@uwaterloo.ca}
\affiliation{Institute for Quantum Computing, University of Waterloo, Ontario, N2L 3G1, Canada}
\affiliation{Department of Physics \& Astronomy, University of Waterloo, Ontario, N2L 3G1, Canada}
\affiliation{Department of Chemistry,  University of Waterloo, Ontario, N2L 3G1, Canada}

\author{P.-N. Roy}
\affiliation{Department of Chemistry, University of Waterloo, Ontario, N2L 3G1, Canada}

\author{R. G. Melko}
\affiliation{Department of Physics \& Astronomy, University of Waterloo, Ontario, N2L 3G1, Canada}
\affiliation{Perimeter Institute for Theoretical Physics, Waterloo, Ontario N2L 2Y5, Canada}

\author{A. Del Maestro}
\affiliation{Department of Physics, University of Vermont, Burlington, VT 05405, USA}

\begin{abstract}
\end{abstract}

\maketitle

{\bf
    Area laws were first discovered by Bekenstein and Hawking~\cite{Bekenstein1973,Hawking1974}, who found that the entropy of a black hole grows proportional to its surface area, and not its volume.  Entropy area laws have since become a fundamental part of modern physics, from the holographic principle in quantum gravity~\cite{tHooft1985,Susskind1995,Bousso2002} to ground state wavefunctions of quantum matter, where entanglement entropy is generically found to obey area law scaling~\cite{Eisert2010}.  As no experiments are currently capable of directly probing the entanglement area law in naturally occurring many-body systems, evidence of its existence  is based on studies of simplified theories~\cite{Bombelli1986,Srednicki1993,Eisert2010}.  Using new exact microscopic numerical simulations of superfluid $^4$He, we demonstrate for the first time an area law scaling of entanglement entropy in a real quantum liquid in three dimensions. We validate the fundamental principles underlying its physical origin, and present an ``entanglement equation of state" showing how it depends on the density of the superfluid.
}

%
\begin{figure}[t]
\begin{center}
\includegraphics[width=0.9\columnwidth]{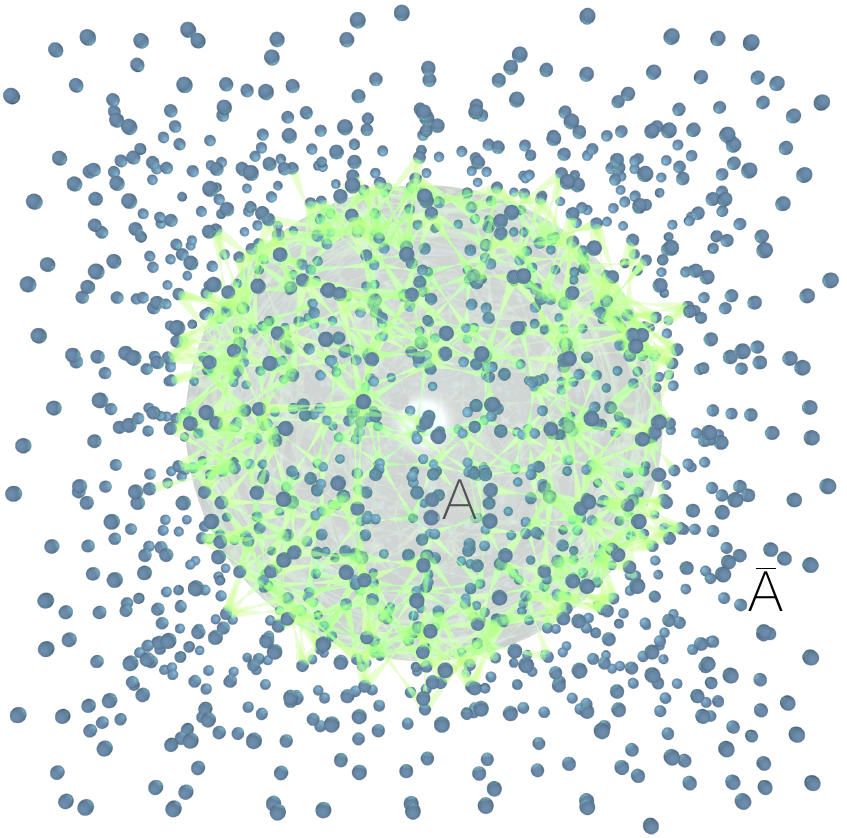}
\end{center}
\caption{\textbf{Entanglement across a spherical boundary.} A container of
    superfluid $^4$He is bipartitioned into a spherical subregion $A$ of radius
$R$ and its compliment $\bar{A}$ at fixed density $n \equiv 1/r_0^3$.  The entanglement between $A$ and $\bar{A}$ is dominated by an area law,
scaling with area of the bounding surface.}
\label{fig:CFT_cor}
\end{figure}
%


Condensed $^4$He undergoes a transition from a normal liquid to a superfluid phase at a critical temperature $T_c \simeq \SI{2.17}{\kelvin}$, at its saturated vapor pressure~\cite{ALLEN1938,KAPITZA1938}.  Superfluid $^4$He was the first experimentally realized, and remains the most extensively studied, quantum phase of matter.  Anomalous phenomena such as dissipationless flow, non-classical rotational inertia, quantized vortices and the Josephson effect, have been thoroughly experimentally characterized \cite{leggett2006quantum}. Early theoretical work demonstrated the quantum mechanical origin of these phenomena~\cite{LONDON1938,TISZA1938,Landau1941,Onsager1949} where the Hamiltonian of liquid $^4$He is that of interacting spinless, nonrelatvistic bosons.  Continuous space quantum Monte Carlo methods enable the precise computation of a wide range of its microscopic and thermodynamic properties, confirming theoretical predictions and reproducing experimental observations~\cite{Ceperley1995}. Moving beyond conventional simulations, recent algorithmic advances have opened up the possibility of measuring entanglement, the non-classical information shared between parts of a quantum state, in numerical experiments \cite{Hastings2010, Herdman2014}.  We combine these two technologies to measure the entanglement entropy in the superfluid phase of bulk $^4$He at zero temperature. Its ground state, $\ket{\Psi}$ in a cubic volume can be bipartitioned into a spherical subregion $A$ and its complement $\bar{A}$ as shown in Figure \ref{fig:CFT_cor}.  To quantify the entanglement between $A$ and $\bar{A}$ in this pure state, we compute the 2nd \ren entanglement entropy 
$S_{2}\left( A \right) \equiv -\log \bigl({ {\rm Tr} \rho_A^{2} }\bigr)$,
where $\rho_A$ is the reduced density matrix of the subsystem: $\rho_A \equiv
\Tr_{\bar{A}} \eket{\Psi}\ebra{\Psi}$.  While other entanglement measures are also of interest, $S_2$ can be measured in numerical simulations as well as experiments on many-body systems without resorting to full state tomography \cite{Islam2015a}.  We investigate its dependence on the radius $R$ of the spherical subregion over a range of densities in the superfluid and find a dominant area law scaling: $S_2 \sim R^2$.


While there is no proof of the area law outside of a restrictive case in one spatial dimension \cite{Hastings:2007bu}, it can be argued to arise from a few fundamental physical principles \cite{Eisert2010,Liu2013,Solodukhin2010,Swingle2010,Grover2011a}. $S_{2}\left( A \right)$ depends on physics local to the entangling surface, and has contributions at all lengthscales $\ell$ ranging from those characterizing microscopic details of the interactions, $r_0$, up to that of the subregion, $R$.  From these, a simple phenomenological scaling theory can be inferred for a spherical boundary of radius $R$ in three dimensions (Figure~\ref{fig:CFT_cor}).  For each infinitesimal region of the bounding surface $d\Sigma$, there is a local contribution to $S_2$ from each lengthscale $\ell$. For a given $\ell$, the lowest order dimensionless quantity that can contribute to $S_2$ is $d\Sigma/\ell^2$; when integrated over the surface this provides a contribution $\sim R^2/\ell^2$. To integrate over lengthscales we must use a logarithmic measure of integration: $S_2 \sim \int_{r_0}^R (R/\ell)^2 d(\log \ell) \sim R^2$; an area law in three dimensions.  While higher order corrections lead to a power series in $R$, the symmetry of the entanglement entropy between complementary regions of pure states, $S(A) = S(\bar{A})$, limits this expansion to even powers (in odd dimensions). This leads to the generic scaling form,
\begin{equation}
S_2 \left( R \right) = 4 \pi a \left({ \frac{R}{r_0} }\right)^2 + b \log
\left({ \frac{R}{r_0} }\right) + c + \mathcal{O} \left(\frac{r_0^2}{R^2}
\right) 
\label{eq:Sscale}
\end{equation}
where  $a$, $b$ and $c$ are dimensionless numbers.  $a$ is non-universal and depends on the microscopic details of the system, while $b$ and $c$ potentially encode universal information that is independent of the short-distance physics.  Note that the leading-order area (as opposed to a volume) scaling and the absence of a subleading linear term, are all features of the underlying physical postulates in three dimensions.

%
\begin{figure}[t]
\includegraphics[width=1.0\columnwidth]{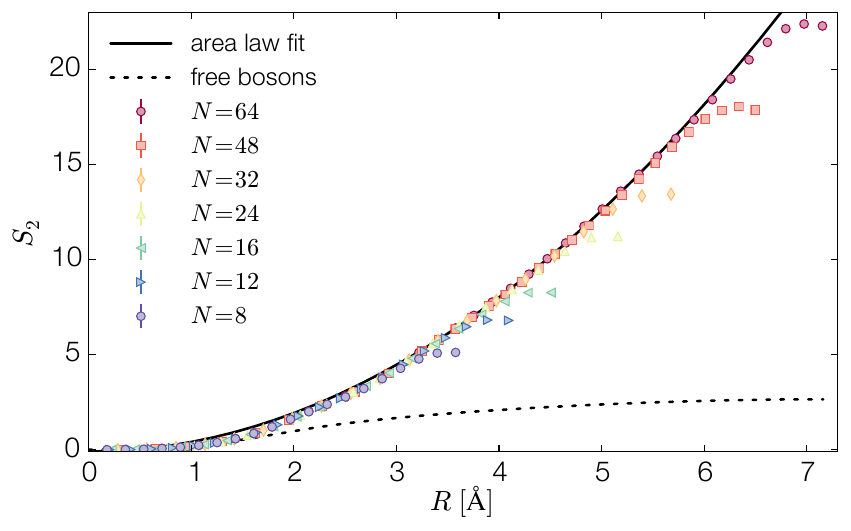}
\caption{{\bf Entanglement entropy of superfluid $^4$He.} The ground state of
$N$ superfluid helium-4 atoms in a cubic cell with linear dimension $L$ and
periodic boundary conditions is bipartitioned into a sphere of radius $R$ and
its complement at fixed equilibrium density $n_0 = N/L^3 \simeq
\SI{0.02186}{\angstrom^{-3}}$. The \ren entanglement entropy $S_2$ in the limit
$R \ll r_0$ is driven by particle fluctuations into the spherical subregion and
is well described by free bosons (dotted line).  For $r_0 \lesssim R \lesssim
L/2$ we perform a two parameter fit $(a,c)$ of the numerical data to the
scaling form in Equation~\eqref{eq:Sscale} with $b=0$, shown as a solid line.  Deviations from the universal curve for each system size occurs as $R$ approaches $L/2$, due to finite-size effects.}
\label{fig:SvRn0}
\end{figure}
%

We perform numerical tests of the general scaling form of Equation~\eqref{eq:Sscale} via high performance simulations of superfluid $^4$He.  Measuring $S_2$ is significantly more computationally complex than for conventional estimators like the energy and required the development of a new algorithm detailed in the Supplementary Information.  The combined results of $S_2(R)$ for $R \le L/2$ are shown in \figref{fig:SvRn0} for the ground state of $^4$He at equilibrium density. 
We find that the entanglement entropy for different numbers of particles $N$
collapses to a nearly universal curve. Before finite size effects dominate near
$R\sim L/2$, and for $R \gtrsim r_0$, we can fit the data to the scaling form
in Equation~\eqref{eq:Sscale} with the dominant behavior captured by a two parameter fit $(a,c)$ shown as a solid line. The extracted value of $a$ is robust within $\sim5\%$ for fits including $b\ne 0$ (details are provided in the Supplementary Information).

%
\begin{figure}[t!]
\begin{center}
\includegraphics[width=1.0\columnwidth]{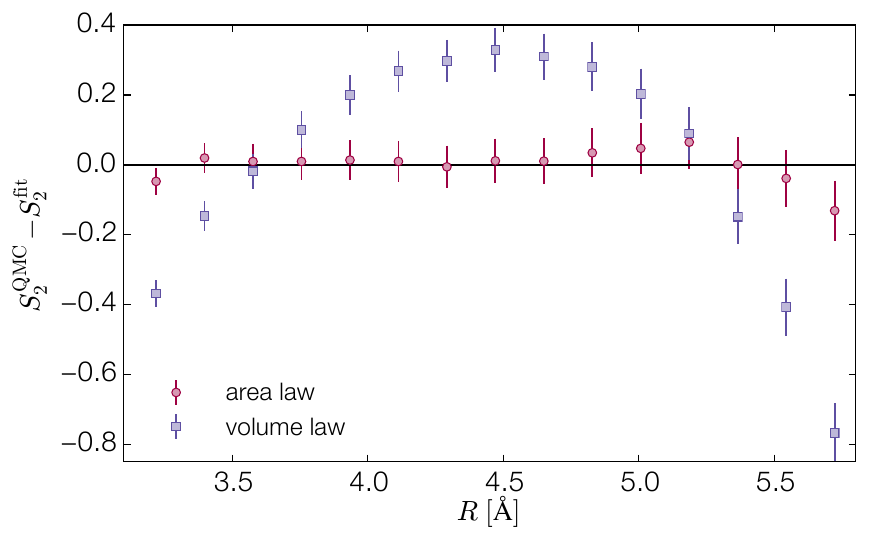}
\end{center}
\caption{{\bf Area or volume law?} The residuals of the \ren entanglement
entropy computed via quantum Monte Carlo (QMC) and two types of two-parameter
fits corresponding to an area law: $S_2^{\rm fit} = 4\pi a (R/r_0)^2 + c$ and a
volume law: $S_2^{\rm fit} = (4\pi a/3) (R/r_0)^3 + c$ for $N=64$ $^4$He atoms
at the equilibrium density $n_0$. Residuals are shown for values of the
spherical subregion radius $R$ over which the fits were performed. The data is
poorly described by the volume law and strongly supports the area law scaling
predicted in Equation~\eqref{eq:Sscale}.}
\label{fig:SvAn0}
\end{figure}
%

The efficacy of this fit and its confirmation of the leading order scaling
behavior of the entanglement is investigated by computing the residuals between
the simulation data and two scaling forms as shown in \figref{fig:SvAn0}.  We
explore the area law predicted by Equation~\eqref{eq:Sscale}, and a volume law
that would be expected for an extensive entropy of thermodynamic origin. The
residuals for the area law are clearly consistent with zero, while strong
deviations for the volume law exclude this as a candidate for the leading order
scaling. The vanishing residuals are consistent with the absence of a
subleading linear correction to the area law, further supporting the physical
postulates leading to Equation~\eqref{eq:Sscale}.  The residuals should display
evidence of any subleading (non-constant) terms commensurate with Equation~\eqref{eq:Sscale}; however, simulations of larger system sizes, providing a wider range of length scales, would be required to reliably study these terms.

\begin{figure}[t]
\begin{center}
\includegraphics[width=1.0\columnwidth]{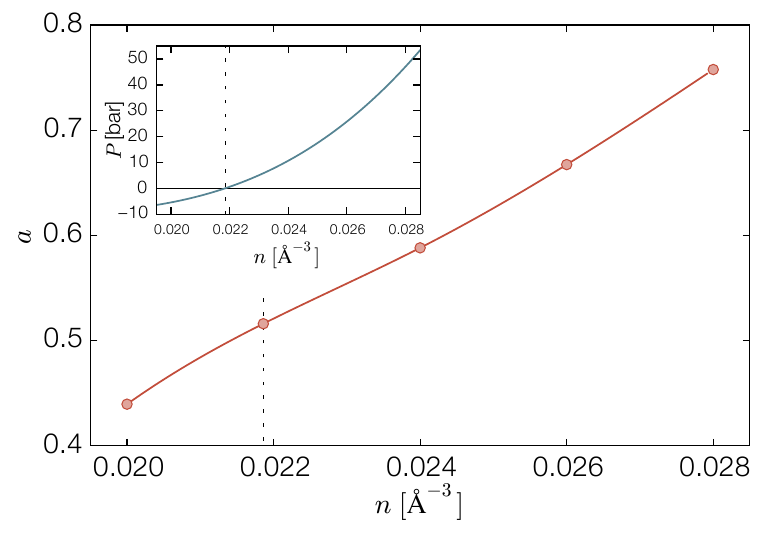}
\end{center}
\caption{{\bf Entanglement equation of state.} The area-law coefficient $a$ vs. density $n$ in the ground state of $^4$He in the superfluid regime, as computed by two parameter fits to quantum Monte Carlo data for $N=64$ (symbols); the line is a guide for the eye. \emph{Inset}: the pressure $P$ of the ground state of superfluid $^4$He as a function of density, from Ref.~\cite{Maris2002}. This suggests that $a$ is a monotonic function of pressure. The dashed vertical lines correspond to the equilibrium density $n_0 \simeq \SI{0.02186}{\angstrom^{-3}}$.}
\label{fig:avn}
\end{figure}

To understand the physical origin of the area law scaling coefficient $a$, we
define an entanglement length scale ${\ell_{\rm{e}} \equiv r_0/\sqrt{a}}$. From
a fit to Equation~\eqref{eq:Sscale} we find $\ell_{\rm{e}} \simeq 1.3 r_0 \approx \SI{5}{\angstrom}$ at $n=n_0$.  This strongly suggests that the short distance physics of the potential hard core and adjacent attractive minima dominate the area law scaling behavior. 
To confirm, we study the effect of the density on the entanglement by computing $S_2$ for superfluid $^4$He over a range of densities near $n_0$ corresponding to positive ($n>n_0$) and negative ($n<n_0$) pressures.  Performing a two-parameter fit to the area law scaling for each density, we plot an ``entanglement equation of state'' in \figref{fig:avn}.  $a$ is an increasing function of density, and thus a monotonic function of pressure (see inset).  We find that $\ell_{\rm{e}}$ depends both on the nature of short distance interaction as well as the interparticle separation. We can contrast this behavior to the non-interacting Bose gas, where $S_2(R)$ is a pure function of the aspect ratio $R/L$. 

%
%

%

In conclusion, we have demonstrated that the prototypical quantum fluid, superfluid $^4$He, displays area law scaling of its entanglement entropy.  Using large-scale, exact microscopic simulations, we have extracted the numerical coefficient of the area law term and find that it is a monotonically increasing function of density.  This confirms that fluctuations and interactions \emph{local} to the entangling boundary drive the physics of the area law. Moreover, this provides quantitative validation of the basic postulates which determine not only the leading-order scaling of the entanglement entropy, but constrains its subleading scaling, some of which may contain new universal physics.  For example, it is predicted that logarithmic corrections should arise due to the existence of a spontaneously broken continuous symmetry in the thermodynamic limit, contributing a universal coefficient due to the presence of a low energy ``tower of states" spectrum and a Goldstone boson~\cite{Song2011a,Kallin2011,Metlitski2011}.  For superfluid $^4$He with a spherical entangling surface, this $b$ coefficient will combine with another universal number arising from the vacuum theory governing the bosonic fluctuations.  This latter quantity encodes one of the two central charges that characterize a three dimensional conformal field theory~\cite{Casini2010}, believed to be the fundamental constant that quantifies how entropy monotonically decreases under renormalization group flow \cite{Cardy1988c}.  A curved bounding surface such as a sphere without defects is only possible in the spatial continuum.  It is thus possible that fundamental physical quantities which arise for smooth geometries are inaccessible in simple lattice models and can only be probed in Galilean invariant quantum liquids.

 \section{Acknowledgments}
We are grateful to L.~Hayward Sierens and M.~Metlitski for useful discussion.
This research was supported in part by the National Science Foundation under Awards No.~DMR-1553991 (A.D.) and PHY-1125915 (R.M.).
Additionally, we acknowledge support from the Natural Sciences and Engineering Research Council of Canada,
the Canada Research Chair program, and the Perimeter Institute for Theoretical Physics.
Computations were performed on the Vermont Advanced Computing Core supported by NASA (NNX-08AO96G) as well as on resources provided by the Shared Hierarchical Academic Research Computing Network (SHARCNET).
Research at Perimeter Institute is supported through Industry Canada and by the Province of Ontario through the Ministry of Research \& Innovation.

\FloatBarrier
\putbib[He4Area]
\end{bibunit}

\widetext
\pagebreak
\begin{center}
\textbf{\large Supplementary Information: Entanglement area law  in superfluid $^4$He}
\end{center}
\setcounter{equation}{0}
\setcounter{figure}{0}
\setcounter{table}{0}
\setcounter{page}{1}
\makeatletter
\renewcommand{\theequation}{S\arabic{equation}}
\renewcommand{\thefigure}{S\arabic{figure}}
\renewcommand{\bibnumfmt}[1]{[S#1]}
\renewcommand{\citenumfont}[1]{S#1}
\begin{bibunit}[apsrev4-1]

\section{Hamiltonian of Helium-4}

The Hamiltonian of bulk liquid $^4$He is that of spinless, nonrelatvistic bosons interacting with a two-body interatomic potential:
\begin{equation}
    H = - \frac{\hbar^2}{2m} \sum_{i=1}^N \nabla_i^2  + \sum_{i<j} V\bigl(\left| \vec{r}_i-\vec{r}_j\right|\bigr). \label{eq:H} 
\end{equation}
Accurate microscopic interatomic potentials $V$ for $^4$He have been developed~\cite{Aziz1979} and their use in conjunction with quantum Monte Carlo methods has allowed for precise calculations of a wide range of ground and finite temperature properties of liquid helium~\cite{Ceperley1995}. The salient features of the interatomic potential $V$ are a repulsive hard core of radius $\sigma\simeq\SI{2.6}{\angstrom}$, an attractive power-law tail, and a minimum of depth $\epsilon \simeq\SI{11}{\kelvin}$ at radius $r_m \simeq \SI{3.0}{\angstrom}$. At the equilibrium number density $n_0\simeq\SI{0.02186}{\angstrom^{-3}}$, the mean inter-particle separation $r_0 \simeq\SI{3.6}{\angstrom}$ is slightly larger than $r_m$.

 \section{Quantum Monte Carlo Method}
 \label{sec:QMC}

In recent work we have developed Monte Carlo methods to compute \ren entropies in systems of itinerant particles in the spatial continuum using path integral Monte Carlo~\cite{Hastings2010,Melko2010,Herdman2014,Herdman2014a}.  Path integral Monte Carlo methods have been successfully used to compute conventional thermodynamic properties of systems of order $N=10^4$ particles.  Due to the increased computational difficulty of computing \ren entropies, the data presented here is limited to systems with $N\le64$ $^4$He atoms. However, systems of this size have been demonstrated to be sufficiently large to display fundamental macroscopic features of superfluid $^4$He~\cite{Ceperley1995}. 

While our previously published algorithm was focused on one spatial dimension, in this section we briefly describe a new variant which allows for the computation of \ren entropies in the ground state of systems of interacting bosons in three dimensional continuous space.  We use a path integral ground state method~\cite{Sarsa2000,Ceperley1995} which gives access to group state properties via imaginary time projection on a trial state $\ket{\Psi_{\rm T}}$:
\begin{equation}
\eket{\Psi} \propto \lim_{\beta \rightarrow \infty} e^{-\beta H} \eket{\Psi_{\rm{T}}}. \label{eq:Psi0_proj} 
\end{equation}
We label the classical configuration space of $N$ bosons by $\R$, which is a vector of particle coordinates. Considering Hamiltonians of the form \Eqref{eq:H}, we use a standard approximation to the imaginary-time propagator 
\begin{equation}
\rho_{\tau} \left( \R, \R'\right) \simeq \ebra{\R} e^{-\tau H} \eket{\R'} 
\label{eq:rho_tau}
\end{equation}
which is accurate to fourth order in the short time $\tau$~\cite{Chin1997a,Jang2001}.
Because \Eqref{eq:rho_tau} is non-negative for bosonic systems, we can Monte Carlo sample discrete imaginary-time world-line configurations of the replicated system, where we use $P$ discrete time steps with $2\beta=P\tau$.

For the 2nd \ren entropy, we define a replicated Hilbert space of two non-interacting copies of the system, $\{\ket{\R}\otimes\ket{\tilde{\R}}\}$. We may compute $S_2$ under a bipartition of the system into $A$ and its complement $\bar{A}$ from the expectation value of a ``swap operator" which swaps the configuration of $A$ between the two replicas:
\begin{align}
\mathrm{SWAP}_A \biggl( \eket{ \R_A,\R_{\bar{A}} } &\otimes  \eket{ \tilde{\R}_A,\tilde{\R}_{\bar{A}} } \biggr) = \eket{ \tilde{\R}_A,\R_{\bar{A}} } \otimes  \eket{ \R_A,\tilde{\R}_{\bar{A}} } ,\notag
\end{align}
where $\R = \{\R_A,\R_{\bar{A}}\}$ such that $\R_{A}$ ($\R_{\bar{A}}$) is a vector of the coordinates of the particles in $A$ ($\bar{A}$). The estimator for $S_2$ is then simply related to the expectation value of the swap operator~\cite{Hastings2010}:
\begin{equation}
S_2 \left(A\right) = -\log \biggl[ \ebra{\Psi}\otimes\ebra{\Psi} \mathrm{SWAP}_A \eket{\Psi}\otimes\eket{\Psi} \biggr]. \label{eq:S2_est}
\end{equation}

\begin{figure}
\centering
  \includegraphics[width=0.45\linewidth]{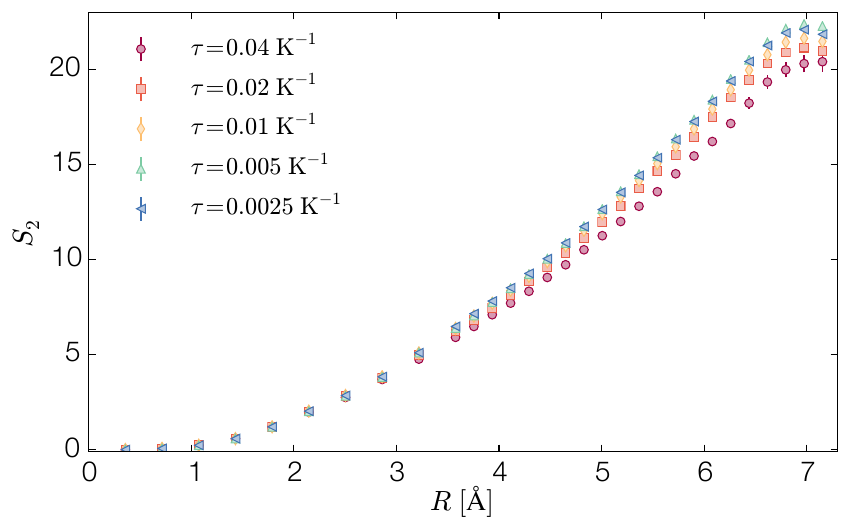}
  \includegraphics[width=0.45\linewidth]{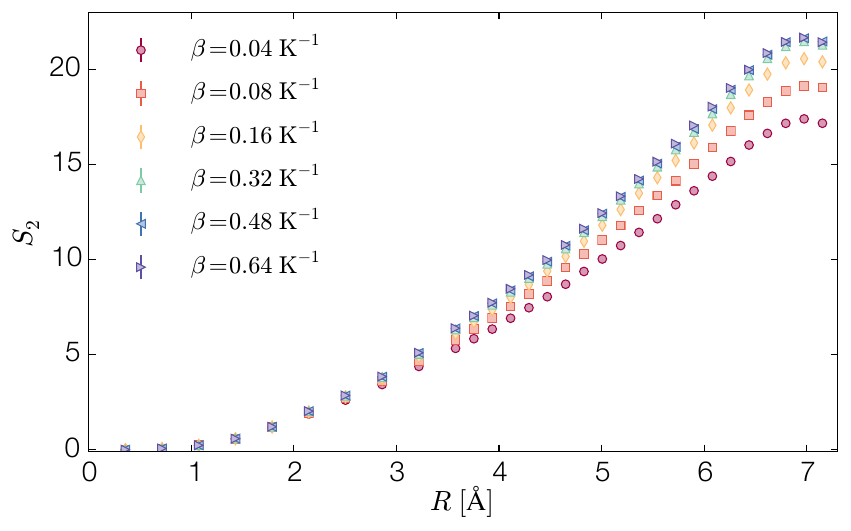}
\caption{Convergence of $S_2$ with (left) discrete imaginary-time step $\tau$ using $\beta=\SI{0.48}{\kelvin^{-1}}$, and (right) with imaginary-time length $\beta$ using $\tau=\SI{0.01}{\kelvin^{-1}}$, for an $N=64$ system at equilibrium density $n_0$.}
\label{fig:SvtP}
\end{figure}

To compute \Eqref{eq:S2_est} with path integral ground state Monte Carlo, we use \Eqref{eq:Psi0_proj}, and consider imaginary time paths of length $2\beta$, capped by $\Psi_{\rm{T}}$ on either end of the path. We Monte Carlo sample an extended configuration space of imaginary time worldlines that includes configurations where worldlines that pass through $A$ at imaginary time $\beta$ may swap between replicas. That is, worldlines that pass through the $\bar{A}$ spatial subregion at time $\beta$ are always propagated in imaginary time along the same replica, but particles that pass through the $A$ subsystem at time $\beta$ may be connected via $\rho_\tau$ to a worldine in $\tilde{\vec{R}}_{A}$ or $\vec{R}_A$ at time $\beta+\tau$. Such swapped worldline configurations have weight of the form
\begin{align}
\Psi_T^* \left(\R_0\right)\Psi_T^* \left(\tilde{\R}_0\right) \prod_{i=0}^{P/2-1}\rho_\tau\left(\R_i,\R_{i+1}\right) \rho_\tau\left(\tilde{\R}_i,\tilde{\R}_{i+1}\right)& \times\notag\\
\rho_\tau^{\bar{A}}\left(\R_{P/2},\R_{P/2+1}\right) \rho_\tau^{\bar{A}}\left(\tilde{\R}_{P/2},\tilde{\R}_{P/2+1}\right) 
&\rho_\tau^A\left(\R_{P/2},\tilde{\R}_{P/2+1}\right) \rho_\tau^A\left(\tilde{\R}_{P/2},\R_{P/2+1}\right)  \times \notag \\
&\prod_{i=P/2+1}^{P}\rho_\tau\left(\R_i,\R_{i+1}\right) \rho_\tau\left(\tilde{\R}_i,\tilde{\R}_{i+1}\right) \Psi_T \left(\R_P\right)\Psi_T \left(\tilde{\R}_P\right) \notag
\end{align}
where $\rho_\tau^{A}$ ($\rho_\tau^{\bar{A}}$) are the reduced propagators for the $A$ ($\bar{A}$) subsystem~\cite{Herdman2014,Herdman2014a} and $\vec{R}_P$ is the configuration of $N$ particles at imaginary time $P\tau$. By including updates that allow for the worldine connectivity to interchange between swapped $(s)$ and unswapped $(u)$ configurations, we may measure $S_2$ from \Eqref{eq:S2_est} from the ratio of the swapped and unswapped generalized partition functions:
\begin{equation}
S_2 = -\log{\frac{Z_s}{Z_u}}. 
\end{equation}
We find this estimator to be more efficient that previous variants for systems above one spatial dimension. To further improve its performance we used a ``ratio method" to build up $A$ from smaller increments~\cite{Melko2010,Herdman2016a}. The systematic errors due to finite $\tau$ and $\beta$, can be made arbitrarily small by increasing $P$ at a computational cost that is polynomial in $P$ and $N$. \figref{fig:SvtP} shows the convergence of $S_2(R)$ with $\tau$ and $\beta$. All results shown in the main text were computed using $\beta=\SI{0.48}{\kelvin^{-1}}$, $\tau=\SI{0.005}{\kelvin^{-1}}$, and a constant trial wave function.\\
%
%

 \section{Data Analysis}
 \label{sec:Data}
%
\begin{figure}[]
\begin{center}
\includegraphics[width=0.45\columnwidth]{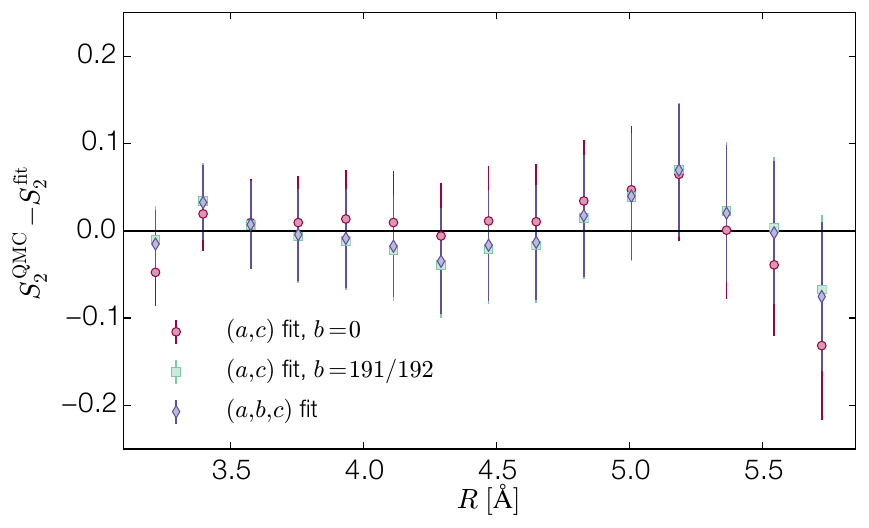}
\end{center}
\caption{Fit residuals for two $(a,c)$ and three $(a,b,c)$ parameters fits QMC data using the scaling form given in Eq. (1) in the main text. All forms are consistent with the data and we find that the area law coefficient $a$ is robust to within $\sim5\%$, independent of the fitting form.}
\label{fig:Sres}
\end{figure}
%
To compute the area law coefficient $a$, we perform fits to the scaling form given in Eq. (1) in the main text. We consider two parameter fits to the QMC data where the subleading logarithmic coefficient $b$ is fixed to zero or its theoretical value, $b=191/192$~\cite{Metlitski2011,Casini2010}, and $(a,c)$ are taken to be free parameters. Additionally we consider a three parameter fit where $(a,b,c)$ are all taken as free parameters. The residuals of these fits relative to the QMC data are shown in \figref{fig:Sres}; we find all three forms are consistent with the QMC data. In all cases we find the area law coefficient is consistent to within $\sim5\%$. Thus we can reliably extract the area law coefficient, independent of the fitting form chosen. To avoid overfitting the data, we extract $a$ using the simplest two parameter form with $b=0$. 

\FloatBarrier
\putbib[He4Area]
\end{bibunit}


\end{document}